\documentclass[a4paper,aps,prb,twocolumn,floatfix]{revtex4}
\usepackage{epsfig}
\usepackage{amsmath}
\usepackage{graphics}
\usepackage{graphicx}
\begin{document}

\title{Co-doped Ceria: Tendency towards ferromagnetism driven by oxygen vacancies}
\author{V. Ferrari $^{1,3,*}$, A. M. Llois$^{1,2,3,*}$, V. Vildosola$^{1,3,*}$}

\affiliation{$^{1}$Departamento de F\'\i sica and INN, Centro At\'omico Constituyentes, Comisi\'on Nacional de Energ\'{\i}a At\'omica, Gral. Paz 1499, 1650 San Mart\'{\i}n, Buenos Aires, Argentina.\\
$^{2}$Departamento de F\'\i sica “Juan Jos\'e Giambiagi”, Facultad de Ciencias Exactas y Naturales, Universidad de Buenos Aires, 1428 Buenos Aires, Argentina.\\
$^{3}$Consejo Nacional de Investigaciones Cient\'\i ficas y T\'ecnicas, C1033AAJ, Buenos Aires, Argentina.\\ 
$^{*}$ These authors contributed equally to this work.}

\date{\today}
\begin{abstract}
We perform an electronic structure study for cerium
oxide homogeneously-doped with cobalt impurities, focusing on the role played by
oxygen vacancies and structural relaxation. By means of full-potential ab-initio methods, we explore the possibility of ferromagnetism
as observed in recent experiments. 
Our results indicate that oxygen vacancies seem to be crucial for the appearance of a ferromagnetic alignment among Co impurities, obtaining an increasing tendency towards ferromagnetism with growing vacancy concentration. The estimated couplings  cannot explain though, the experimentally observed room-temperature ferromagnetism. 
In this systematic study, we draw relevant conclusions regarding the location of the oxygen vacancies and the magnetic couplings involved. In particular, we find that oxygen vacancies tend to nucleate in the neighborhood of Co impurities and we get  a remarkably strong
ferromagnetic coupling between Co atoms and the Ce$^{3+}$ neighboring ions. The calculated magnetic moments per cell depend on the degree of reduction which could explain the wide spread in the magnetization values observed in the experiments.

\end{abstract}


\maketitle

\section{Introduction}

The terrain of the so called Diluted Magnetic Oxides (DMO) is currently
being explored with a strong  drive  to find  Room Temperature (RT)
ferromagnetic (FM) materials for  possible technological applications \cite{DietlReview}.  DMO
are experimentally obtained  by doping the oxide matrix with a small amount
of magnetic Transition Metal (TM) ions. This procedure may introduce
ferromagnetism in otherwise non magnetic materials and opens up
the feasibility of applications that range from spintronics to magneto-
optical devices.

Since the pioneering works of Dietl \textit{et al.} \cite{Dietl}
in Mn-doped ZnO, RT ferromagnetism was observed in various doped oxide
hosts \cite{ChambersReview06} with different TM ions and doping concentrations. 
Some DMO are FM insulators while others are semiconductors and, so
far, there is not an obvious link between the ferromagnetic behavior
and the conduction properties. The samples are sensitive both to preparation
methods and growth conditions and this observation supports the idea
that ferromagnetism might be closely linked to defects and/or to the
presence of oxygen vacancies. Interesting for spintronic applications
are  critical temperatures being far above
room temperature.

Among DMO, Co-doped CeO$_{2}$ (Co$_{x}$Ce$_{1-x}$O$_2$) has attracted particular interest
due to ferromagnetic behavior observed well above RT for
low Co-doping concentration \cite{paperC,EtgensAPL07,paperA,paperH,Wen}.
Moreover, ceria is a transparent and high dielectric constant rare-earth
oxide, whose fluorite structure matches well with that of silicon. It keeps
the crystal structure, both, under doping and upon the formation of oxygen
vacancies, thus promising a good integrability for spintronic devices
even for non-stoichiometric compounds.  However, there is controversy among the experimental results
regarding critical temperatures and the values of the magnetic moments, as a function
of doping concentration. As a matter of fact, a wide range of magnetic moments has been reported \cite{paperA,EtgensAPL07,paperC,Wen}
and this  has raised concerns about the intrinsic nature of the
FM properties of these materials due, for example, to Co secondary phases, heterogeneities or even contamination.

The presence of oxygen vacancies has been considered as a possible factor affecting
the FM response in the case of films \cite{paperH} but it is not yet clear
whether it induces Co clustering and/or promotes magnetic ordering. It has been recently argued that even without introducing magnetic impurities, the presence of oxygen vacancies in CeO$_{2}$ could stabilize ferromagnetism \cite{superexchangeceria}.

There is experimental evidence available, obtained using several techniques \cite{Henderson,Liu,mullins1,Esch,EtgensJPCM08}, indicating that the charge left behind by the oxygen vacancies gets localized near some Ce atoms driving  the Ce$^{4+}$ to Ce$^{3+}$ reduction. In the case of undoped Ceria surfaces, this localization has been recently addressed by Ganduglia-Pirovano {\it et al.}\cite{VGP_PRL}. The charge localization happens both in doped as well as in  undoped  CeO$_{2-\delta}$, and has to be properly taken
into account since it might affect atomic relaxation and, in turn, the
magnetic behavior. It is important to remark that in order to perform a theoretical study of the magnetic
properties of Co-doped reduced ceria, it is critical to relax the crystal structure.

Previous ab-initio DFT+U calculations have reported  a detailed description
of the relaxation processes in reduced ceria  upon doping with  transition metal atoms such as  Zr\cite{paperE} and
Pd\cite{paperD}. It results that relaxation is different in Zr-doped systems as compared to the Pd-doped
ones, because the electronic distribution, which changes upon reduction \cite{paperD}, depends on the size and the chemical characteristics of the dopant. 

Our goal in this paper is to explore the role played
by oxygen vacancies on the magnetic properties of a bulk ceria matrix as a function of Co-doping concentration. We perform Local Density Approximation (LSDA)+U calculations\cite{Wien2k} considering
bulk unit cells with substitutional replacement of Co atoms into cerium sites \cite{bulknote}.  
In this work we take into account,  in the relaxation process,  the presence of Ce$^{3+}$ ions induced by reduction. We also consider the evolution of the magnetic interactions with the number
of oxygen vacancies per Co atom, which (to the best of our knowledge) has not yet been  reported for doped ceria.

We organize the paper as follows: In Sec. \ref{sec:Method-of-calculation} and Sec. \ref{sec:supercell}, the computational method is described. In Sec. \ref{sec:Location-vacancies} and Sec.  \ref{sec:Ionic-relaxation}, the effect of introducing oxygen vacancies in Co-doped ceria is studied. In Sec. \ref{sec:properties}, the  magnetic and electronic properties as a function of the dopant concentration and oxygen deficiency  are explained  and finally, in Sec. \ref{sec:conclusions}, the results  of this work are discussed.

\section{Method of calculation\label{sec:Method-of-calculation}}

Density functional theory (DFT) \cite{DFT} calculations are performed within the Local Density Approximation (LDA) \cite{LDA}. We use the full potential augmented plane waves method as implemented in the Wien2k code \cite{Wien2k}, where the space is
divided into muffin tin (MT) spheres around the atoms and an interstitial region. Plane waves
are used to describe the region outside the spheres. The muffin tin
radii used are R$_{MT}^{Co}$ =1.9 au, R$_{MT}^{O}$ =1.6 au, R$_{MT}^{Ce}$
=2.3 au. The number of plane waves in the interstitial region is set by the cut-off  parameter RK$_{max}$=7 (or 6 for the biggest cells) where R is the minimum  R$_{MT}$ in the corresponding cell. To attain the desired convergence precision, we use a 7 X 7 X 7 k-mesh in the Brillouin zone for a Co-doping concentration of x=12.5 \% and a 5 X 5 X 5 mesh for x=6.25  \%.

In the cases where we have charge localization on the 4$f$ states of Ce (which we call Ce$^{3+}$), the structures are first allowed to relax taking these 4$f$ electrons as core levels \cite{note}, until forces on atoms are below 1 mRy/au. Then, the  electronic structure of the fully relaxed supercell is obtained using the LSDA+U approach  switching on the local Coulomb interaction in the Ce$^{3+}$ ions \cite{LDAplusU} with  U$_{eff}$=U - J = 6 eV \cite{Ueffceria}. 
The relaxation processes involved will be described in detail in Sec. \ref{sec:Ionic-relaxation}.

\section{Supercell calculations\label{sec:supercell}}

The crystal structure of CeO$_{2}$ is fluorite with an experimental
lattice parameter of 5.411 \AA . Each Ce$^{4+}$ cation is coordinated
to eight O$^{2-}$ nearest neighbors and in turn, each O$^{2-}$ is tetrahedrally
coordinated to Ce$^{4+}$ cations.

Previous theoretical results for doped ceria indicate very small changes
in the lattice constants upon doping with different transition metals,
while preserving the cubic symmetry \cite{paperE,paperD}. In particular, when doping with Zr, a 0.2\% 
contraction in the lattice parameters is obtained \cite{paperE} and a 0.04\% reduction, when doping with Pd \cite{paperD}. Taking this fact into account, we consider the experimental lattice constants of unreduced CeO$_{2}$ for
all the systems under study. However, as we are interested in understanding the effect of both, magnetic impurities and  oxygen vacancies, it is essential to relax the internal atomic positions, as mentioned in the previous section.

\begin{figure}[ht]
\begin{centering}
\includegraphics[scale=0.7]{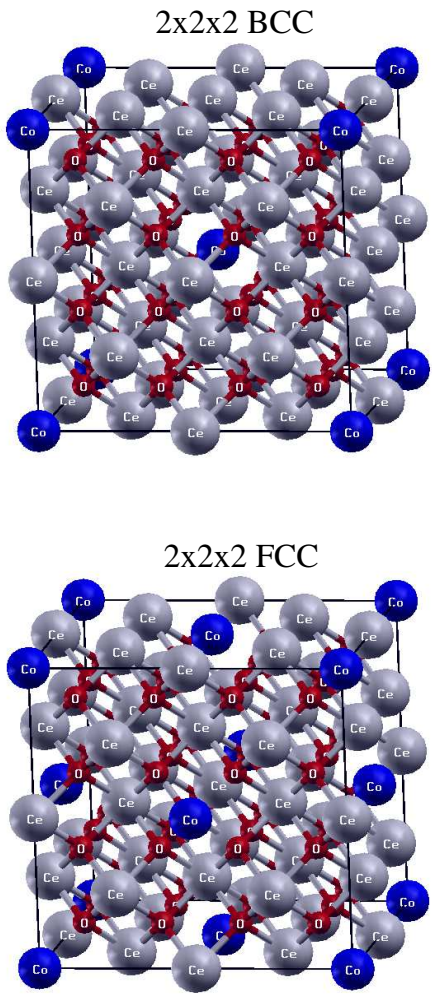} 
\par\end{centering}
\caption{(Color on-line) SuperCells considered for two different Co-doping concentrations. 
Top: 2x2x2 BCC type supercell in which one of the 16 cerium sites
is replaced by a Co atom. This supercell contains 48 atoms.
Bottom: 2x2x2 FCC supercell where one of the 8 cerium sites is replaced
by a Co atom. This is a 24 atom-supercell. \label{supercell}}
\end{figure}

For a concentration of Co given by  x=6.25 \%, we use a 2x2x2 BCC-type supercell, and for x=12.5 \%, a 2x2x2 FCC-type one. The supercells are built out of the conventional 12-atoms cubic unit cell of CeO$_{2}$ and are  schematically shown in Fig. \ref{supercell}. For the unreduced
compounds these supercells contain 48 and 24 atoms, respectively.
The nearest neighbor distance between Co atoms is 9.37\AA~ in the BCC cell and 7.65 \AA~
in the FCC one, so that these atoms  can be safely  considered as impurities.

For each concentration we study configurations with zero, one or two
vacancies per Co impurity atom.
When an oxygen vacancy is created in pure bulk CeO$_{2}$, two electrons
are left in the system and the vacancy is surrounded by four next nearest neighbor Ce atoms. As mentioned before, several experimental
results seem to indicate that these extra electrons occupy localized
$4f$ states on two Ce sites. There are then, two types of
Ce atoms: Ce$^{4+}$ ions in a 4$f^0$-like configuration and Ce$^{3+}$ ions in a 4$f^1$ one, that need to be considered theoretically (see for instance Ref. \onlinecite{Skorodumova1}
and \onlinecite{Review-Ganduglia}, and references therein).

When Co impurities are introduced, the system gains in complexity. X-ray experiments suggest
that in  oxygen deficient samples the valence of Co is 2+ (see Refs. \onlinecite{EtgensJPCM08,paperH,Wen}).
Regarding Ce, the presence of Ce$^{3+}$ has been reported in Refs.
\onlinecite{EtgensJPCM08, paperH}. In particular, in Ref. \onlinecite{EtgensJPCM08},
for a sample with a 4.5 \% Co concentration, the results seem
to indicate that around 10\% of the Ce atoms are Ce$^{3+}$ while the others are Ce$^{4+}$.

Taking the above into account, in this work we consider a
variety of situations, depending on the number of oxygen vacancies
per impurity Co atom,  as it will be detailed in the following section.

\section{Location of the oxygen vacancies\label{sec:Location-vacancies}}

\begin{figure}[htb]
\begin{centering}
\includegraphics[scale=0.7]{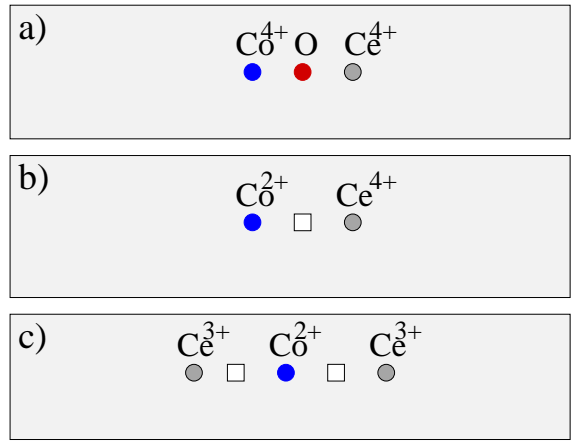} 
\label{squeme-vac}
\par\end{centering}
\caption{(Color on line) Schemes of the lowest energy  configurations of oxygen vacancies within the ceria matrix for: (a) Unreduced system. (b) One vacancy per Co atom. (c) Two vacancies per Co. The white square indicates an oxygen vacancy.}
\label{Flo:scheme-vac} 
\end{figure}

The first question which arises when introducing vacancies into the Co-CeO$_{2}$ system relates to their location within the ceria matrix:  what is energetically more favorable, to have oxygen vacancies close to or far from the Co impurities? To answer this question we first place one vacancy in a  site nearest neighbor to a Co atom and relax the structure for the case x=12.5 \%.  We then put the oxygen vacancy far away from the Co atom and after relaxing the structure, compare both results. 
In view of the experimental evidence mentioned in the previous section and the model suggested by Vodungbo {\it et al.} \cite{EtgensAPL07}, we consider that the extra electrons left behind by the vacancy,  migrate to the Co atoms leaving all the Ce ions as Ce$^{4+}$, for both vacancy locations.  The difference in energy between the two considered vacancy configurations is 0.55 eV per Co atom 
in favor of the vacancy being close to the impurity (see Fig \ref{Flo:scheme-vac}.(b).

We also consider a larger concentration of vacancies, namely, two per Co impurity atom and  perform the calculations for x=6.25 \%.
We take into account two possibilities: a configuration where the two oxygen vacancies are close to the Co atom as suggested in Ref. \onlinecite{EtgensAPL07} (see scheme in Fig. \ref{Flo:scheme-vac}.(c))
and a configuration where one oxygen vacancy is close to the impurity  while the other is far away from it. Comparing the total energies of both configurations, we find that the two oxygen vacancies prefer to be close to the Co atom by 0.62 eV per Co atom.

It is clear from these results that cobalt has a strong tendency to nucleate oxygen vacancies in ceria. In the following sections we  discuss the relaxation process, the electronic structure and the magnetic couplings for the most
favorable spacial distribution of vacancies.

\section{Ionic relaxation \label{sec:Ionic-relaxation}}

\subsection{Zero oxygen vacancies}

As discussed in the previous section, for the doped system without oxygen vacancies, both Co and Ce  are expected to be +4  and  therefore we use  conventional LSDA calculations to relax the  supercells for the two impurity concentrations considered. The obtained relaxations are qualitatively similar: The eight oxygen atoms which surround  the impurity get symmetrically closer to it by 0.09 \AA~ (for x=6.25\%) and by 0.07 \AA~ (for x=12.5\%). Regarding the nearest cerium atoms, they move towards the impurity by around 0.04 \AA~ for x=6.25\%,  while they do not relax due to symmetry reasons for the largest concentration considered.

\subsection{One oxygen vacancy per Co atom}

The valence configuration is depicted  in Fig. \ref{Flo:scheme-vac}.(b). As  there are no localized 4$f$ electrons in the cell for this vacancy concentration, we use again conventional LSDA calculations and focus only on x=12.5\%.

After relaxation, the seven oxygen atoms, which stay  around the impurity after creating the vacancy, undergo a displacement of around  0.06 \AA~ in average: one oxygen moves towards Co considerably ($\sim$0.3 \AA), three oxygen atoms  get closer to it by $\sim$0.11\AA~ and the last three ones move outwards  by $\sim$0.07 \AA. Regarding the nearest neighbors of the vacancy, the nearest three Ce ions and the Co impurity move outwards with respect to it (the Ce atoms displace by 0.1 \AA~ and the Co atom by 0.28 \AA) while the six nearest neighbor oxygen atoms move towards the vacancy by  0.15 \AA \, on the average.

\subsection{Two oxygen vacancies per Co atom}

We consider the most stable vacancy location, namely, the two vacancies being near the impurity atom. As mentioned above, in this case, the extra electronic charge left behind by the vacancy,  localizes at  two Ce atoms, which are  nearest neighbors to Co.

After relaxation, the six oxygen atoms surrounding the transition metal impurity move towards it.  For x=6.25 \%, two of them displace by $\sim$0.27 \AA~  while  the other four do it by $\sim$0.15 \AA. The oxygen atoms which are nearest neighbors to the  Ce$^{3+}$ ions move away from them in average about 0.1 \AA. This expansion of all oxygen atoms around the Ce$^{3+}$ has to do with the coulomb repulsion due to the extra charge localized at that Ce site. 
In contrast, this is not observed around the  Ce$^{4+}$ ions, where some of the nearest neighbor oxygen atoms get closer to these ions while others move away, but without undergoing a net expansion of the oxygen-cloud.  Qualitatively similar atomic relaxations are obtained for  x=12.5\%.
It is clear that the introduction of vacancies produces an important rearrangement of the structure.

\subsection{Comparison with other dopants}

It is interesting to compare the results obtained when doping with Co with those appearing in the literature for other dopants such as  Zr or Pd. The available data include cases considering just one vacancy per impurity \cite{paperE,paperD}.

For unreduced ceria, relaxation in the presence of Zr  impurities is similar to
that of Co: the 8 nearest neighbor oxygen atoms and the three nearest  Ce$^{4+}$ ions  move towards
the impurity by $\sim$0.1 \AA~ and $\sim$0.036 \AA~, respectively \cite{paperE}. This is due to the fact that Zr and Co
have smaller atomic radii than the  Ce$^{4+}$ ion.  When doping with Pd, which has an atomic radius  similar to  the ionic radius of Ce$^{4+}$, the displacements towards the impurity  are smaller than in the case of Zr and Co \cite{paperD}.

In the presence of one oxygen vacancy, relaxation for Co-doping resembles neither the Pd nor the Zr dopant cases.  Co and Zr move away from the vacancy while Pd moves towards it. All nearest neighbor oxygen atoms get closer to the vacancy when the dopants are Co and Pd. In the case of Zr, most of those oxygen ions move towards the vacancy except for one oxygen that moves away from it  \cite{paperE}. 

The above mentioned differences in the relaxation processes might be due to different reasons: In the case of Zr-doping, the charge left behind   by the vacancy localizes on two neighboring Ce atoms, while for Pd and Co it localizes both at the impurity site and at its nearest neighbor oxygen atoms.  When doping with Pd, the difference in relaxation might be due to the different size of the dopant atomic radii.

\section{Magnetic and electronic properties of ceria  as a function
of Co doping and reduction. \label{sec:properties}}

\subsection{Magnetic coupling between cobalt and $Ce^{3+}$ ions \label{sec:coupling-Co-Ce}}

\begin{figure}[h]
\begin{centering}
\includegraphics[scale=0.7]{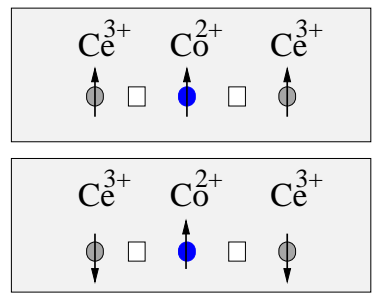} 
\label{squeme-vac}
\par\end{centering}
\caption{(Color on line) Scheme of the two configurations considered for calculating the magnetic coupling between Co and Ce$^{3+}$ ions (Top, FM. Bottom, AFM). The FM configuration is favoured by an energy difference of 120meV per Co atom.}
\label{fig:JCoCe} 
\end{figure}

Before going into the analysis of the evolution of the electronic and magnetic properties of ceria as a function of dopant and  vacancy concentration, it is interesting to determine  the nature of the magnetic coupling between Ce$^{3+}$ and Co.   From the above discussion, it is clear that this coupling takes place  when two vacancies per Co are present.   To get insight into this problem, we consider the case of x=12.5\%  to compare the  energies of FM and AFM spin-alignments  of  Ce$^{3+}$ with respect to Co, as depicted in Fig. \ref{fig:JCoCe}.
The energy difference between these configurations is 120 meV per Co atom in favor of the FM alignment. 
Cobalt and Ce$^{3+}$ ions show, then,  a strong ferromagnetic coupling.

\subsection{Electronic structure and magnetic coupling among dopants \label{sec:electronicstructure}}

\begin{figure}[htb]
\begin{centering}
\includegraphics[clip,scale=0.25]{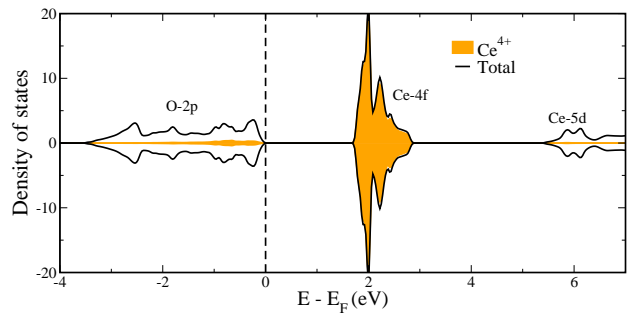}
\par\end{centering}
\caption{(Color on line) Total density of states for pure CeO$_{2}$ (black line) and partial one
for Ce (orange). \label{fig:DOS-ceria}}
\end{figure}

Pure CeO$_{2}$, without  impurities or oxygen vacancies, is a non-magnetic insulator. The occupied bands are  mainly of O \emph{2p} character and the unoccupied ones are mostly of Ce \emph{4f} character (See Fig. \ref{fig:DOS-ceria}). When impurities are introduced in the unreduced system, defect states appear due to  majority and minority contributions coming from the $d$ states  of the substitutional Co atoms.  For both impurity concentrations, and no oxygen vacancies,  the magnetic coupling among Co ions is slightly antiferromagnetic. In Table \ref{tab:Energies}, we show the  energy differences  among FM and AFM configurations, $\Delta$ E =E$_{FM}$ -E$_{AF}$ for both Co-concentrations and different number of vacancies per Co.

\begin{table}[htb]
\begin{center}\begin{tabular}{|c|c|c|}
\hline
case & $\Delta$E (x=6.25\%)  & $\Delta$E (x=12.5\%)   \\
\hline
\hline
0 vac & +0.4  & +2.5 \\
\hline
1 vac   & 0.0  & -0.5    \\
\hline
2 vac   & -3.1   &   -12.5  \\
\hline
\end{tabular}\end{center}
\caption{$\Delta$E =E$_{FM}$ - E$_{AF}$, energy difference  per Co ion  (in meV) between FM and AF solutions, to estimate the magnetic coupling between Co atoms, for different cobalt concentration x and number of vacancies.}
\label{tab:Energies}
\end{table}

 When oxygen vacancies are introduced, a tendency towards FM Co-Co coupling appears. The energy difference $\Delta$ E changes sign thus favoring a growing FM alignment, which increases with, both, vacancy and Co concentration. For the largest Co concentration, the vacancy-induced ferromagnetic Co-Co coupling, is clearly established.

 Experimental evidence seems to suggest that there are no metallic Co clusters or secondary phase formation in Co-doped ceria \cite{paperA,paperC}. In this work, assuming then, that Co impurities are diluted and homogeneously distributed within the ceria matrix, a tendency  towards ferromagnetism is obtained. This tendency increases with oxygen deficiency. However, the energy differences obtained cannot explain the high Curie temperatures observed in the experiments. Thus, the underlying mechanism for the observed room temperature ferromagnetism, remains unclear.

\begin{figure}[htb]
\begin{centering}
\includegraphics[clip,scale=0.6]{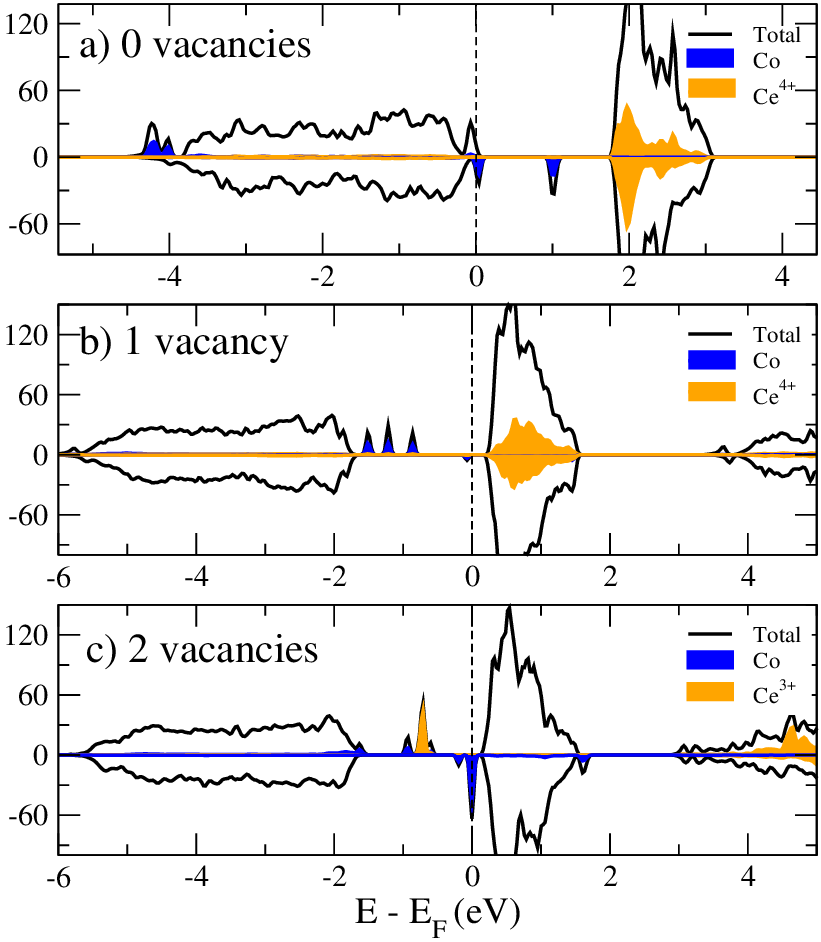} 
\par\end{centering}
\caption{(Color on line) Density of states for $Ce_{1-x}Co_{x}O_{2-\delta}$ with x=6.25\% for (a) unreduced system ($\delta$=0), (b) one vacancy per Co ($\delta$=0.0313), and  (c) two vacancies per Co ($\delta$=0.0625).   Total DOS (black line), partial DOS for cobalt (blue) and cerium  (orange). In (a) and (b), there is no charge localization on any Ce site. In (c) the partial DOS of a Ce$^{3+}$ is shown. \label{fig:DOS-BCC}}
\end{figure}

\begin{figure}[htb]
\begin{centering}
\includegraphics[clip,scale=0.6]{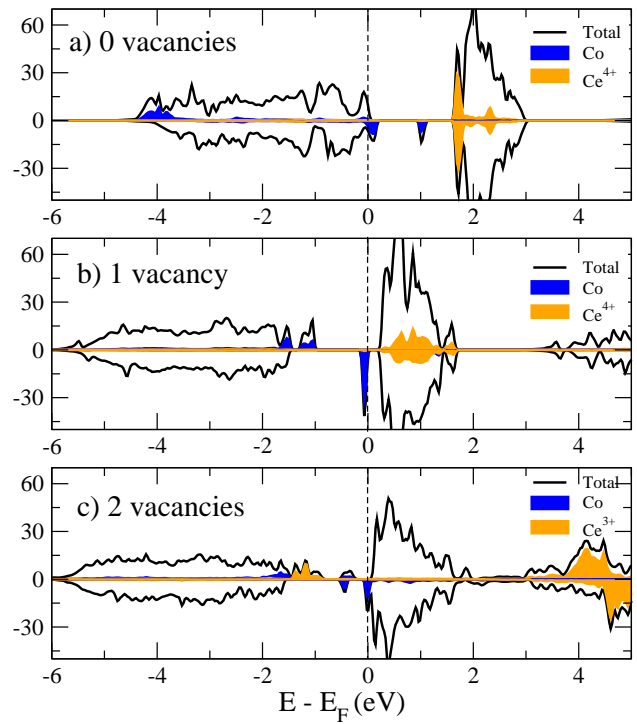} 
\par\end{centering}
\caption{(Color on line) Density of states for $Ce_{1-x}Co_{x}O_{2-\delta}$ with x=12.5\% for (a) unreduced system ($\delta$=0), (b) one vacancy per Co ($\delta$=0.0625), and  (c) two vacancies per Co ($\delta$=0.125).  Total DOS (black line), partial DOS for cobalt (blue) and cerium  (orange). In (a) and (b),
there is no charge localization on any Ce site. In (c) the partial
DOS of a Ce$^{3+}$ is shown. \label{fig:DOS-FCC}}
\end{figure}

To compare the effect that the introduction of vacancies has on the electronic structure of doped ceria,  we show the densities of states corresponding to the FM solutions.
As it can be observed in Fig. \ref{fig:DOS-BCC}.(a) and Fig. \ref{fig:DOS-FCC}.(a), for the two unreduced Co concentrations considered here, two broad impurity peaks develop at the bottom of the majority valence band of O-$p$ character. 
The most important contribution to the peak appearing at the top of the majority valence band, comes from the  oxygen atoms which are nearest neighbors of the  Co impurities.  The orbitals of these oxygen atoms are magnetically polarized due to hybridization  with the Co $d$ impurity orbitals.    
Two localized minority impurity peaks  lie in the gap of ceria.  For both Co concentrations, the Fermi level falls  within the first minority peak, which lies  at the top of the oxygen minority $p$ band.  The second and unoccupied  minority  impurity peak  appears well inside  the original gap of ceria, far from the band edges.

\begin{figure}[htb]
\begin{centering}
\includegraphics[clip,scale=0.6]{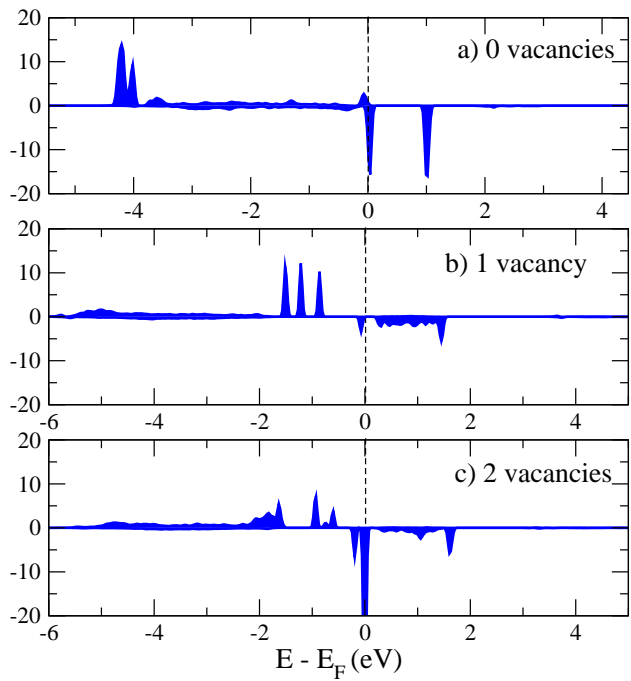}
\par\end{centering}
\caption{Local density of states on the Co impurity site, with increasing oxygen vacancy concentration for Co concentration x=6.25\%. (a) without vacancies, (b) one vacancy per Co, (c) two vacancies per Co. \label{figdosCo-BCC}}
\end{figure}

In order to better follow  the evolution of the impurity peaks with the introduction of vacancies, we show in Fig. \ref{figdosCo-BCC} the  projected local densities of states on Co for x=6.25\%.  The trends are similar for x=12.5 \%,  with more delocalized impurity bands due to  larger Co-Co hybridization.

When  oxygen vacancies are introduced,  Co$^{4+}$  reduces to Co$^{2+}$.  In Fig. \ref{fig:DOS-BCC}.(b) and (c)  and Fig. \ref{fig:DOS-FCC}.(b) and (c), the corresponding  total densities of states and the Co local densities of states are shown.  Due to the Coulomb repulsion  the large  spectral  weight of Co  $d$ character, originally lying  at the botton of the majority valence band goes  into  two new majority peaks,  which can be observed above this  band.  There are now three majority Co impurity peaks lying below the Fermi level and above the valence band in these  reduced configurations for the two  Co concentrations considered.

When  two oxygen vacancies per Co  are present, the extra charge, which  localizes on two Ce$^{3+}$ sites, gives rise to the  occupied majority  peak of  \emph{4f} character, which can be  observed in Fig. \ref{fig:DOS-BCC}.(c) and Fig. \ref{fig:DOS-FCC}.(c).

\vspace{2cm}

\subsection{Evolution of the magnetic moments\label{sec:magneticmoments}}

In the non reduced impurity systems, the  total magnetic moment of the FM cells is mainly due to the polarization of Co and its nearest neighbor  oxygen atoms.  The AFM solution  shows a  distribution of the magnetic polarization among the atoms which   is not different from the one obtained for the FM solution.  The  hybridization effects are, namely,  very local. The eight nearest neighbor oxygen atoms  of Co  become ferromagnetically polarized with respect to the impurity, for both Co Concentrations.  See Tables \ref{tab:Magnetic-moments} and \ref{tab:Magnetic-moments-1}. 

When we introduce one vacancy, the two extra electrons which are  left behind give rise to a decrease in the total magnetic moment of the ferromagnetic  cell.  The total moment goes  from 4.76 $\mu_{B}$ (for x=6.25\%) and  4.60$\mu_{B}$ (for x=12.5\%)  to 3 $\mu_{B}$.  This decrease in the value of the  magnetic moment per Co atom of almost 2 $\mu_{B}$  points towards an effective Co valence which goes from +4 to +2. The extra charge left behind by the oxygen vacancy localizes, then,  in the electronic cloud surrounding the transition metal ion, which is covalently shared by this ion and the seven remaining nearest neighbor oxygen atoms.  

Finally, when there are two vacancies per Co, 
the magnetic moment per magnetic ion  is   5 $\mu_{B}$,  for both Co concentrations.   From the  four extra electrons per Co atom left behind by the vacancies, two   contribute  to decrease the magnetic moment of the cloud formed by  Co and its neighboring
oxygen atoms with respect to the unreduced situation. The other two electrons  get localized on  two Ce$^{3+}$  sites close
to the vacancies, giving rise  to  the increase in 2 $\mu_{B}$  of the magnetic moment of the cell  per Co atom which goes from 3 $\mu_{B}$ to 5$\mu_{B}$. See Tables \ref{tab:Magnetic-moments} and \ref{tab:Magnetic-moments-1}. 

The values of the total  magnetic moments of the ferromagnetic cells also reveal that the introduction of  vacancies changes the band character of the doped systems.  The unreduced impurity systems show a non integer  cell magnetic moment, characteristic of  a metallic-like band structure, while the introduction of vacancies gives rise to  half-metallic-like features, as for instance the integer value of the total  cell spin moments.

It is interesting to remark, again, that the  total magnetic moment per cell strongly depends  on  the degree of reduction. This might  explain the wide spread  in the magnetization values obtained in the experiments as reported in the literature \cite{paperA,EtgensAPL07,paperC,Wen}.

\begin{table}[htb]
\begin{center}\begin{tabular}{|c|c|c|c|c|c|}
\hline 
case & $\mu_{T}$ & $\mu_{Co}$ & $\mu_{O}^{nn}$ & $\mu(Co+O^{nn})$ & $\mu(Ce')$\\
\hline 
\hline
0 vac  & 4.76  & 2.94  & 1.28 & 4.22  & 0.04 \\
\hline 
1 vac  & 3.00  & 2.50  & 0.42   & 2.92  &-0.17 \\
\hline 
2 vac  & 5.00  & 2.49  & 0.37   & 2.86  & 0.95 \\
\hline
\end{tabular}\end{center}

\caption{Magnetic moments (in $\mu_{B}$) for x=6.25 \%: $\mu_{O}^{nn}$ means the sum of the
magnetic moments of all oxygen atoms nearest neighbor to cobalt. $\mu(Co+O^{nn})$
means the magnetic moment of the cloud formed by the Co and its nearest
neighbors. $\mu(Ce')$ is the one corresponding to a cerium atom close
to the vacancy. For the 2 vacancies case, there are two reduced Ce per cobalt. \label{tab:Magnetic-moments}}

\end{table}

\begin{table}[htb]
\begin{center} \begin{tabular}{|c|c|c|c|c|c|}
\hline 
case & $\mu_{T}$ & $\mu_{Co}$ & $\mu_{O}^{nn}$ & $\mu(Co+O^{nn})$ & $\mu(Ce')$\\
\hline 
\hline
0 vac  & 4.60   & 2.90   & 1.28   & 4.18  & 0.04 \\
\hline 
1 vac  & 3.00   & 2.50   & 0.40   & 2.90  &-0.03 \\
\hline 
2 vac  & 5.00   & 2.41   & 0.40   & 2.81  & 0.99 \\
\hline
\end{tabular}\end{center}
\caption{Same caption as Table \ref{tab:Magnetic-moments}, for x=12.5 \%. \label{tab:Magnetic-moments-1}}
\end{table}

\section{Discussion and conclusions}\label{sec:conclusions}

In this work we have studied the effect of oxygen vacancies on the magnetic and electronic properties of Co-doped ceria.
We have shown that  the introduction of oxygen vacancies is essential to drive  a  ferromagnetic coupling among the dopant impurities. Actually, in the absence of vacancies there is a slight tendency towards an antiferromagnetic coupling that changes towards a ferromagnetic one, when the vacancy concentration increases.

For two vacancies per Co, part of  the left-behind charge localizes in  Ce ions sitting in the vicinity of the impurity, turning them into magnetic Ce$^{3+}$ ions.  We find that the  magnetic  coupling among the Co impurities and the Ce$^{3+}$ ions, is strongly ferromagnetic.  This coupling  provides  a  large magnetic moment,  localized  close to the impurity.

Even if Co impurities in unreduced ceria show large local magnetic moments, the interaction among impurities is small for the two dopant concentrations  studied.  The same can be said about the systems with only one vacancy per impurity atom. 
The ferromagnetic coupling among impurities in the presence of two vacancies per Co increases when going from an impurity concentration of  6.25 \% to 12.5 \%.

We conclude that,  although there is a clear tendency towards ferromagnetism in Co doped ceria when oxygen vacancies are present, the obtained FM couplings cannot  explain the high Curie temperatures observed in the experiments.
 On the other hand, the large values of the magnetic moments per Co atom observed  can be understood if the Co impurities, the vacancies and the nearest Ce$^{3+}$ ions build complexes of the type considered in this work.  To understand the simultaneous appearance of the two phenomena, namely Curie temperatures above RT and a wide spread  in the large values of the magnetic moments per Co dopant, it might be necessary to consider eventual inhomogeneities and/or to introduce other magnetic interaction mechanisms  which are not present in the  exchange interactions taken into account within the framework  of these DFT calculations.

\section*{ACKNOWLEDGMENTS}

We are thankful to V. Etgens and F. Vidal who motivated these calculations and gave us experimental insight.
We  acknowledge useful discussions with R. Weht, J. Milano and M. Weissmann. In particular, we are indebted to R. Weht regarding  the calculation of the magnetic coupling for the supercell-BCC system.  This work was funded by CONICET, ANPCyT and UBA (Argentina) through grants PIP-CONICET-6016, UBACYT-X123, PICT06-157, PICT06-1765 and PICT05-33304.


\begin{thebibliography}{28}
\bibitem[{bul()}]{bulknote}
\bibinfo{note}{We consider bulk systems taking into account that the film
  thickness in the experiments is at the micrometer scale (See Refs.
  \onlinecite{EtgensAPL07,paperH,Wen,paperGconference}).}
\bibitem[{not()}]{note}
\bibinfo{note}{We have checked the relaxation procedure with the 4$f$ electrons
  of Cerium treated as core levels by relaxing the structure of undoped and
  reduced CeO$_{2-\delta}$ and comparing the results with the literature
  \cite{paperD,Fabris}. We obtained a quantitative agreement regarding both the
  ionic relaxation and the eletronic structure.}
\end{thebibliography}

\end{document}